\documentclass[prl, twocolumn, superscriptaddress, reprint]{revtex4-1}

\usepackage{amsmath} 
\usepackage{amsfonts}
\usepackage{amssymb}
\usepackage{graphicx}  
\usepackage{hyperref}  
\usepackage{color}

\begin{document}
\bibliographystyle{apsrev4-1}

\title{Weaving knotted vector fields with tunable helicity}
\author{Hridesh Kedia}
\email{hridesh@uchicago.edu}
\affiliation{Department of Physics,  James Franck Institute, Enrico Fermi Institute, The University of Chicago, 929 E 57th St, Chicago, IL 60637, USA}

\author{David Foster}
\affiliation{HH Wills Physics Laboratory, University of Bristol, Tyndall Avenue, Bristol BS8 1TL, UK}

\author{Mark R.~Dennis}
\affiliation{HH Wills Physics Laboratory, University of Bristol, Tyndall Avenue, Bristol BS8 1TL, UK}

\author{William T.M. Irvine}
\affiliation{Department of Physics,  James Franck Institute, Enrico Fermi Institute, The University of Chicago, 929 E 57th St, Chicago, IL 60637, USA}

\begin{abstract}
We present a general construction of divergence-free knotted vector fields from complex scalar fields, whose closed field lines encode many kinds of knots and links, including torus knots, their cables, the figure-8 knot and its generalizations.
As finite-energy physical fields they represent initial states for fields such as the magnetic field in a plasma, or the vorticity field in a fluid. 
We give a systematic procedure for calculating the vector potential, starting from complex scalar functions with knotted zero filaments, thus enabling an explicit computation of the helicity of these knotted fields. 
The construction can be used to generate isolated knotted flux tubes, filled by knots encoded in the lines of the vector field. 
Lastly we give examples of manifestly knotted vector fields with vanishing helicity. 
Our results provide building blocks for analytical models and simulations alike.
\end{abstract}

\maketitle

{\it Introduction.}
The idea that a physical field---such as a magnetic field---could be weaved into a knotty texture, has fascinated scientists ever since Lord Kelvin conjectured that atoms were in fact vortex knots in the aether. 
Since then, topology has emerged as a key organizing principle in physics, and knottiness is explored as a fundamental aspect of physical fields including classical and quantum fluids \cite{moffatt_degree_1969, enciso_knots_2012, kleckner_creation_2013, scheeler_helicity_2014, barenghi_knots_2007, kawaguchi_knots_2008, hall_tying_2016, kleckner_how_2016}, magnetic fields in light and plasmas \cite{kamchatnov_topological_1982, moffatt_magnetostatic_1985, ranada_topological_1989, chui_energy_1995, irvine_linked_2008, irvine_linked_2010, arrayas_class_2015, thompson_constructing_2014, smiet_self-organizing_2015, ranada_ball_2000, kedia_tying_2013}, liquid crystals \cite{tkalec_reconfigurable_2011, alexander_colloquium:_2012, machon_knots_2013, martinez_mutually_2014}, optical fields \cite{berry_phase_2000, dennis_isolated_2010}, nonlinear field theories \cite{faddeev_stable_1997, battye_solitons_1999, babaev_hidden_2002, sutcliffe_knots_2007}, and superconductors \cite{babaev_dual_2002, babaev_non-meissner_2009}.

In particular, helicity---a measure of average linking of field lines---is a conserved quantity in ideal fluids \cite{helmholtz_uber_1858,thomson_vortex_1869} and plasmas \cite{woltjer_theorem_1958,chandrasekhar_force-free_1958,newcomb_motion_1958}. 
Helicity thus places a fundamental topological constraint on their evolution \cite{moffatt_degree_1969, moffatt_magnetostatic_1985}, and is known to play an important role in turbulent dynamo theory \cite{monchaux_von_2009,steenbeck_berechnung_2014,moffatt_helicity_2014}, magnetic relaxation in plasmas \cite{arnold_asymptotic_1974, freedman_note_1988}, and turbulence \cite{rogers_helicity_1987, wallace_experimental_1992}.
Beyond fluids and plasmas, helicity conservation leads to a natural connection between the minimum energy configurations of knotted magnetic flux tubes \cite{moffatt_magnetostatic_1985,freedman_note_1988,moffatt_energy_1990}, and tight knot configurations \cite{katritch_geometry_1996,pieranski_ideal_2001} which have wide-ranging applications in polymers \cite{bajer_preface_2013} and molecular biology \cite{stasiak_topological_2013}, and tentatively with the spectrum of mass-energies of glueballs in the quark-gluon plasma \cite{buniy_model_2003,buniy_glueballs_2005,buniy_tight_2014}.

The difficulty of constructing knotted field configurations explicitly, and controlling their helicity, makes it challenging to understand the role of helicity in the evolution of knotted structures \cite{moffatt_degree_1969,moffatt_magnetostatic_1985,chui_energy_1995}.
Tying a knot in the lines of a vector field is a more subtle affair than tying a shoelace into a knot: all the streamlines of the entire space-filling field must twist to conform to the knotted region.

Here, we show how to construct divergence-free, finite-energy vector fields which are arbitrarily knotted with tunable helicity explicitly.
Furthermore, we give a systematic prescription for calculating the helicity of these knotted fields. 
These fields should be useful in providing a better understanding of the interplay between topology and dynamics, especially for magnetic fields in plasmas or vorticity fields in fluid flows.

\begin{figure}[!htb]
\includegraphics[width = \columnwidth]{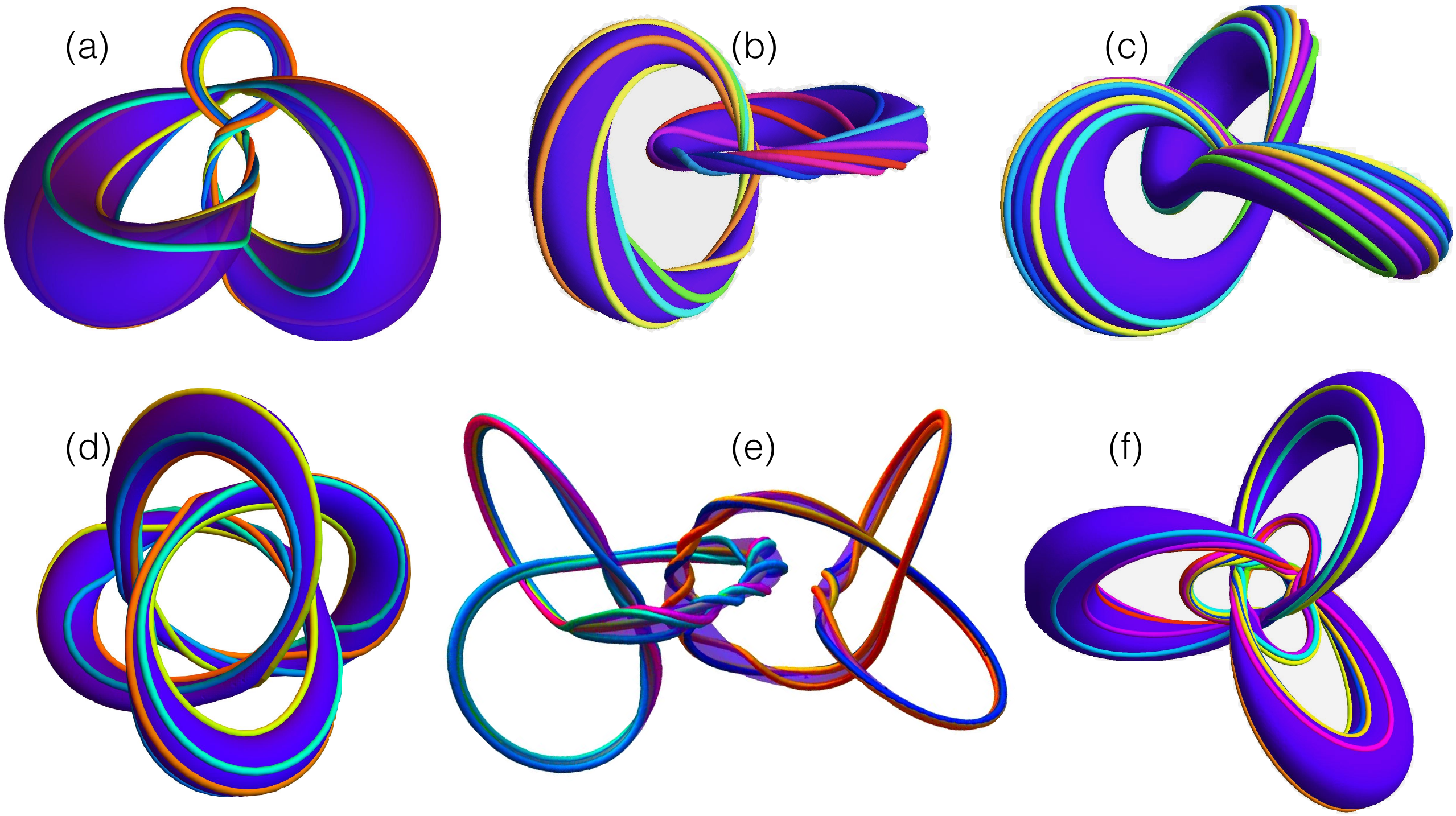}
\caption{
   Knotted structures encoded in the level sets of the complex scalar fields $\psi = P(u,u^\ast, v, v^\ast)/Q(u,u^\ast, v, v^\ast)$, where $(u,v)$ are defined in Eq.~(\ref{uv_r3}). 
   ({\bf a}) Figure-8 knots: ${\small \psi=}$ ${\small v/\left(64u^3-12u(3-2 v^2+2 {v^\ast}^2)+(14 v^2 + 14 {v^\ast}^2 - v^4 + {v^\ast}^4)\right)}$. 
   ({\bf b}) Linked rings: ${\small \psi=u^2/\left( u^2 - v^2 \right)}$. 
   ({\bf c}) Trefoil knots: ${\small \psi= u^3/\left(u^3+v^2\right)}$. Level curves of $\psi$ encode torus knots and links when $Q(u, v)$ is of Brieskorn form \cite{milnor_singular_1969}: $u^p+v^q$.
   ({\bf d}) Figure-8 knots (symmetric): ${\small \psi=}$ ${\small u/\left(64 v^3 -12 v(3+2u^2-2{u^\ast}^2 )-( 14 u^2 + 14 {u^\ast}^2 + u^4 - {u^\ast}^4 )\right)}$.
   ({\bf e}) Linked trefoil knots, constructed from 2 copies of the Milnor polynomial for a trefoil knot. See Supplemental Material for details. 
   ({\bf f}) $C^{2,3}_{3,2}$ cable knots: ${\small \psi=}$ ${\small \left(u\,v\right)/\left(v^4-2u^3\,v^2-2iu^3\,v + u^6 +\frac{1}{4}u^3\right)}$. 
   }
\label{fig:psi_fibres}
\end{figure}%

A classical problem from mathematics is the study of knots and links as nodal lines (zeros) of complex scalar fields \cite{brauner_verhalten_1928, milnor_singular_1969, perron_noeud_1982, dennis_isolated_2010, taylor_geometry_2014}.
In fact, the level sets of a complex scalar field can give rise to collections of knotted curves that smoothly intertwine to fill up space. 
Well-known examples are the Hopf fibration \cite{ranada_topological_1989,lyons_elementary_2003,urbantke_hopf_2003,irvine_linked_2008, irvine_linked_2010,mosseri_hopf_2012}, Seifert fibrations \cite{sadoc_3-sphere_2009,arrayas_class_2015} and Milnor fibrations \cite{milnor_singular_1969,dennis_isolated_2010,king_knotting_2010}.

Some representative examples of such complex scalar fields are illustrated in Fig.~\ref{fig:psi_fibres}, where the level curves wind around knotted or linked tori, encoding the Hopf link (Fig.~\ref{fig:psi_fibres}(b)), the trefoil knot (Fig.~\ref{fig:psi_fibres}(c)), and 
further generalizations including the figure-8 knot \cite{dennis_isolated_2010} (Fig.~\ref{fig:psi_fibres}(a),(d)), links of knots (Fig.~\ref{fig:psi_fibres}(e)) and cable knots (Fig.~\ref{fig:psi_fibres}(f)). 
In all of these examples, the level curves of the complex scalar field $\psi$, for any complex value of $\psi$, organize around a core set of lines corresponding to lines where $\psi = 0$ and $\infty$. 
Our construction of knotted vector fields follows from such knotted complex scalar fields, based on \cite{brauner_verhalten_1928,milnor_singular_1969,sutcliffe_knots_2007}, where the level curves of constant complex amplitude are collections of knotted curves filling up space. 

A vector field tangent to the level curves of a complex scalar field $\psi$ is given simply by the the cross product $-i\nabla\psi^\ast\times\nabla\psi = \nabla\times\mathrm{Im}\left(\psi^* \nabla \psi\right)$.
A vector field with the same flow lines is 
\begin{equation}
\mathbf{B}=\frac{1}{2\pi i}\frac{\nabla\psi^\ast\times\nabla\psi}{\left( 1+\psi\,\psi^\ast\right)^2}. \label{kfc}
\end{equation}
This field is smooth everywhere, divergence-free ($\nabla\cdot\mathbf{B} = 0$) and has finite energy ($\int \mathrm{d}^3x \, |\mathbf{B}|^2 < \infty$).
This vector field arises in a variety of different contexts, and was used previously to construct knotted initial states for electromagnetic fields \cite{ranada_topological_1989,arrayas_class_2015}, and topological solitons in ideal magnetohydrodynamics \cite{kamchatnov_topological_1982}.

Since the flow lines of $\mathbf{B}$ (i.e.~the level sets of $\psi$) can clearly be knotted, it is natural to suppose that such fields have nontrivial helicity. 
Explicitly calculating the helicity $\mathcal{H}=\int {\rm d}^3x\, \mathbf{A}\cdot\mathbf{B}$ requires the choice of a vector potential $\mathbf{A}$ such that $\nabla\times\mathbf{A}=\mathbf{B}$.
A natural candidate,
\begin{equation}
\mathbf{A}=\frac{1}{4 \pi i}\frac{ \left( \psi^\ast \, \nabla\psi - \psi \, \nabla\psi^\ast \right)}{(1+\psi^\ast\,\psi)}, \label{A_nat}
\end{equation}
may tempt one to infer that the helicity of the knotted vector field $\mathbf{B}$ vanishes. 
We will show that $\mathbf{A}$ in Eq.~(\ref{A_nat}) has a singular part which can be systematically removed, leading to a nonsingular vector potential which allows explicit calculation of the helicity of all these knotted fields.

Put together, we can construct knotted vector fields which encode a wide variety of topologies including torus knots and links, the figure-8 knot, cable knots, and arbitrarily linked combinations of these knots. 
The helicity of these fields can be computed explicitly, and may be varied without changing the underlying knotted structure.
Furthermore, these fields may be restricted to the interior of knotted flux tubes, whose helicity can be calculated exactly.
We also show how to construct knotted fields with vanishing total helicity, but possessing non-trivial helicity in a subregion of space.

\begin{figure}[!hbt]
\includegraphics[width = \columnwidth]{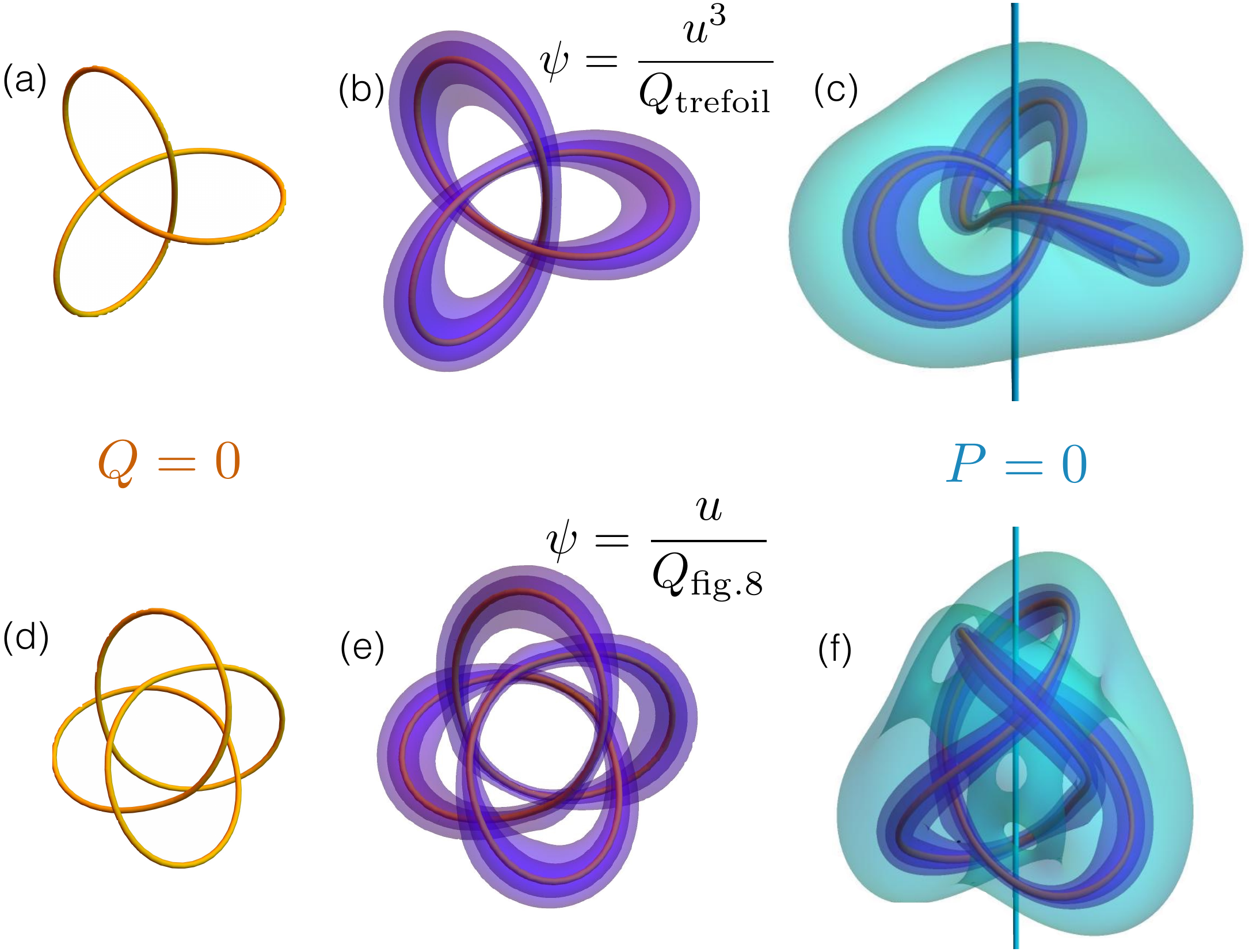}
\caption{
   Organization of the lines of $\mathbf{B}$ around lines where $\psi=P(u,v)/Q(u,v)$ is $0$ or $\infty$. 
   ({\bf a}), ({\bf d}) $Q=0$ corresponds to the trefoil and figure-8 knots. 
   $Q_{\rm trefoil}=u^3+v^2$\,,\,$Q_{\operatorname{fig-8}}=64 v^3 -12 v(3+2u^2-2{u^\ast}^2 )-( 14 u^2 + 14 {u^\ast}^2 + u^4 - {u^\ast}^4)$.
   ({\bf b}),({\bf e}) The lines of $\mathbf{B}$ are tangent to nested knotted tori (blue) organized around the knots where $Q=0$. 
   ({\bf c}),({\bf f}) $P(u,v)=0$ corresponds to the $z$-axis. 
   The lines of $\mathbf{B}$ are tangent to nested tori (cyan) organized around $P(u,v)=0$.}
\label{rational_map}
\end{figure}%

{\it Rational maps.}
Rational maps have found success in approximating certain minimum energy solutions of the Skyrme model \cite{houghton_rational_1998}, and this technique was extended by Sutcliffe \cite{sutcliffe_knots_2007} to approximate knotted solutions of the Skyrme-Faddeev model.
The knotted vector field construction described here is based on rational maps of similar form.
A rational map is defined as the ratio of two complex-valued polynomials $\psi = P(u,u^\ast,v,v^\ast)/Q(u,u^\ast,v,v^\ast)$, where the nodal lines (zeros) of $Q(u,u^\ast,v,v^\ast)$ have the form of the desired knot, and $P(u,u^\ast,v,v^\ast)$ is chosen to encode the desired helicity. 
Here, as in Fig.~\ref{fig:psi_fibres}, $(u,v)$ are complex coordinates on ${S}^3$ which stereographically project to coordinates $(x,y,z)$ in $\mathbb{R}^3$ by 
\begin{equation}
u = \frac{2(x+ i\, y)}{1+r^2}\,,\, v = \frac{2z+i\,(r^2-1)}{1+r^2}, \label{uv_r3}
\end{equation}
where $r^2=x^2+y^2+z^2$, and $(u^\ast,v^\ast)$ denote complex conjugates of $(u,v)$.

Such $\psi$ automatically give rise to a vector field $\mathbf{B}$ as in Eq.~(\ref{kfc}), whose flow lines coincide with the level curves of $\psi$. 
The core set of lines that organize the level curves of $\psi$ are the zeros of $P$ and $Q$ (see Fig.~\ref{rational_map}). 
A wide range of knots can be encoded in the zeros of $Q(u,u^\ast, v, v^\ast)$,  including $(p,q)$-torus knots $(Q = u^q + v^p)$, the figure-8 knot \cite{dennis_isolated_2010} and various generalizations \cite{king_knotting_2010}, implying that a wide variety of knotted fields can be constructed using rational maps.

{\it Structure of knotted field lines}.
To explicate the structure of the lines of $\mathbf{B}$, we rewrite Eq.~(\ref{kfc}) in terms of Euler potentials \cite{stern_geomagnetic_1967, stern_euler_1970, hesse_theoretical_1988, khurana_euler_1997}:
\begin{align}
\mathbf{B} &=\nabla\left( \frac{\psi\,\psi^\ast}{1+\psi\,\psi^\ast} \right)\times \frac{1}{4\pi i}\nabla\log\left( \frac{\psi}{\psi^\ast} \right)   \nonumber \\
&= \nabla\chi\times \nabla\eta \label{kfc_euler_pot} 
\end{align}
where $\chi =\left( \psi\,\psi^\ast \right)/\left(1+\psi\,\psi^\ast \right)$, $\chi\in [0,1]$  and $\eta = \tfrac{1}{4\pi i}\log\left(\psi/\psi^\ast \right)$, $\eta\in [0,2\pi)$. 
We note that as $r\to\infty$, $|\nabla\chi |\sim O(1/r)$, $|\nabla\eta |\sim O(1/r)$, so that the energy density $|\mathbf{B}|^2 \sim O(1/r^4)$, so the energy of all such fields, as the square integral of $\mathbf{B}$, is finite.

\begin{figure}[!hbt]
\includegraphics[width = \columnwidth]{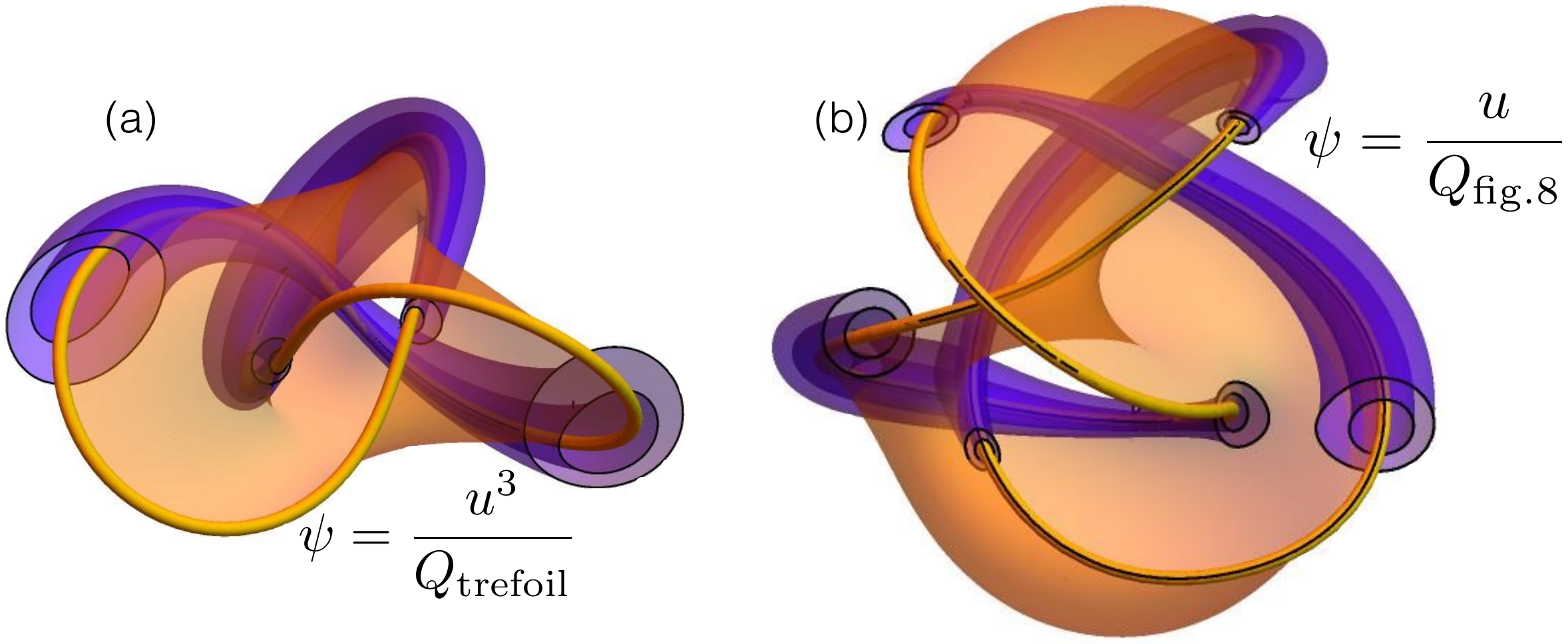}
\caption{
   Knotted field structures: knotted flux surfaces (blue) are surfaces of constant $\chi$, and Seifert surfaces for $Q(u,u^\ast,v,v^\ast)=0$ (orange) are surfaces of constant $f = \left(1/4\pi i\right)\log\left(Q/Q^\ast\right)$.
   ({\bf a}) Trefoil knot with $Q_{\rm trefoil}$, ({\bf b}) Figure-8 knot with $Q_{\operatorname{fig-8}}$ are defined in Fig.~\ref{rational_map}.
   }
   \label{fig:seifert_flux}
\end{figure}

The lines of $\mathbf{B}$ are tangent to surfaces of constant $\chi$ (see Fig.~\ref{fig:seifert_flux}) and surfaces of constant $\eta$ (Seifert surfaces), which can be considered as a generalization of the surfaces of constant $\rho$ and constant $\phi$ in cylindrical coordinates $(\rho,\phi,z)$, with the knot $Q=0$ replacing the $z$-axis. 

The surfaces of constant $\chi$ are knotted tori, nested inside one another, with smaller values of $\chi$ corresponding to larger tori. 
The largest value, $\chi=1$, corresponds to the knot $Q=0$ (see Fig.~\ref{rational_map}(a),(d)), and smaller values of $\chi$ correspond to the nested tori enclosing the knot as shown in Fig.~\ref{fig:seifert_flux}. 
As $\chi$ decreases these knotted tori grow larger, eventually colliding to give tori organized around $P=0$, as shown in Fig.~\ref{rational_map}(c),(f) in cyan. 
These nested tori converge to the lines $P=0$ as $\chi\to 0$.

By contrast, the surfaces of constant $\eta$ are Seifert surfaces for the core set of lines: $P=0\,,Q=0$. 
Seifert surfaces for $Q=0$ are shown in  Fig.~\ref{fig:seifert_flux}.
Since $\phi$ is well-defined only in a multiply-connected volume which excludes the core set of lines, the helicity of $\mathbf{B}$ can be non-vanishing \cite{rosner_relationship_1989}, in spite of being expressible in terms of Euler potentials.

{\it Helicity calculation.}
In order to explicitly calculate the helicity of these knotted fields, it is necessary to define a vector potential $\mathbf{A}$ which is smooth, and satisfies $\nabla\times\mathbf{A}=\mathbf{B}$. 
We now give a general prescription for computing such a vector potential, starting by rewriting $\mathbf{A}$ in Eq.~(\ref{A_nat}) as
\begin{equation}
\mathbf{A}=\frac{1}{4\pi i}\left(\frac{\psi\,\psi^\ast}{1+\psi\,\psi^\ast}\right)\nabla\log\left(\frac{\psi}{\psi^\ast}\right).  \label{A_nat_pole}
\end{equation}
Substituting $\psi=P(u,u^\ast,v,v^\ast)/Q(u,u^\ast,v,v^\ast)$ gives
\begin{eqnarray}
&&\mathbf{A} =\frac{1}{4\pi i}\times \label{A_rational_pole} \\
&&\left[\frac{\vert P\vert^{2}\,\nabla\log\left(\frac{P}{P^\ast}\right) + \vert Q\vert^{2}\,\nabla\log\left(\frac{Q}{Q^\ast}\right)}{\vert P\vert^{2} + \vert Q\vert^{2} }- \nabla\log\left(\frac{Q}{Q^\ast}\right)\right] \nonumber  
\end{eqnarray}
The last term containing $\nabla\log\left(Q/Q^\ast\right)$ is singular at $Q=0$.
Since $\vert Q\vert^{2}\,\nabla\log(Q/Q^\ast) = Q^\ast \,\nabla Q - Q \,\nabla Q^\ast$, this term in the fraction is smooth and nonsingular.

Therefore the singular gauge transformation $\tilde{\mathbf{A}} = \mathbf{A} + \nabla f$, where $f$ is the multivalued function $f = \left(1/4\pi i\right)\log\left(Q/Q^\ast\right)$, removes the singularity in $\mathbf{A}$, allowing the helicity to be computed directly.
The vector potential $\tilde{\mathbf{A}}$ is smooth everywhere, giving the correct helicity $\mathcal{H} = \int {\rm d}^3x\, \tilde{\mathbf{A}}\cdot\mathbf{B}$, which by the Whitehead integral formula is equal to the Hopf invariant of the map $\psi$ \cite{whitehead_expression_1947,sutcliffe_knots_2007}.
Hence we can explicitly compute the helicity~\footnote{Mathematically, $\psi(u,v)$ takes its value on the complex projective plane, $\mathbb{CP}^1$ (i.e.~the complex numbers with the point at $\infty$), which is homeomorphic to the 2-sphere $S^2$: in this sense, helicity can be understood as the topological degree of the map $S^3 \to S^2$.} of arbitrary knotted fields $\mathbf{B}$ and therefore, the Hopf invariant of arbitrary rational maps.

Surfaces of constant $f$ yield explicit expressions for Seifert surfaces of the knot $Q(u,u^\ast,v,v^\ast)=0$ (see Fig.~\ref{fig:seifert_flux}), and could be used to generate initial wave-functions describing knotted vortices in superfluids and Bose-Einstein condensates.

The simplest illustration of our construction arises in the case of the Hopf map \cite{ranada_topological_1989,lyons_elementary_2003,urbantke_hopf_2003,irvine_linked_2008,mosseri_hopf_2012} $\psi = u/v$.
The vector potential given by Eq.~(\ref{A_nat_pole}) has a singularity at $v=0$ (the unit circle in the $xy$-plane), which is removed via the singular gauge transformation $\tilde{\mathbf{A}} = \mathbf{A} + \nabla f$, where $f = \left( 1/4\pi i\right)\log\left(v/v^\ast\right)$. 
The new vector potential $\tilde{\mathbf{A}}$ is smooth everywhere, and gives the correct helicity $\mathcal{H} = \int {\rm d}^3x\, \tilde{\mathbf{A}}\cdot\mathbf{B} = 1$, equal to the Hopf invariant of the map \cite{whitehead_expression_1947,sutcliffe_knots_2007}.
 \begin{figure*}[!htb]
\includegraphics[width = 2\columnwidth]{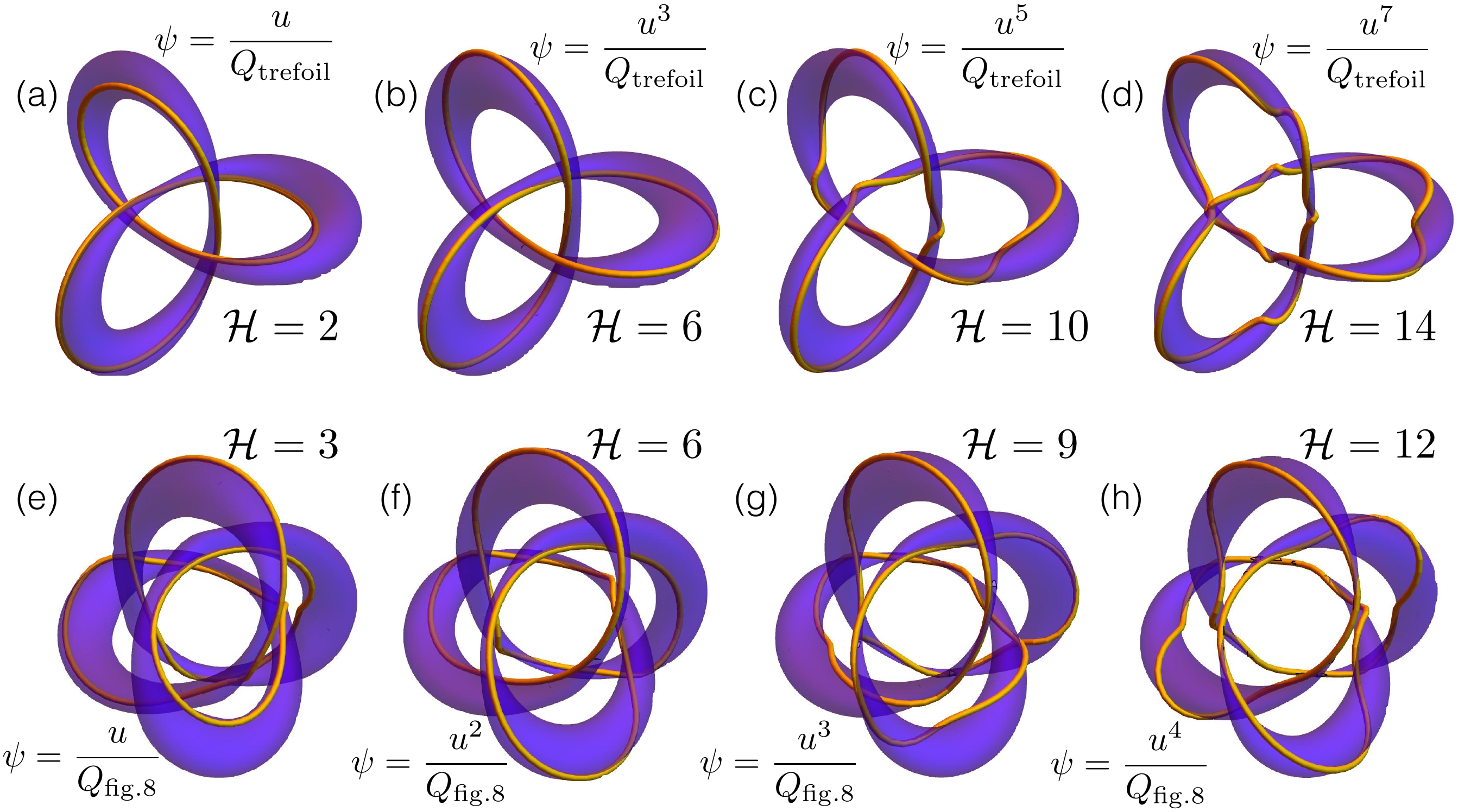}
\caption{
   Tuning the helicity $\mathcal{H}$ of the knotted field $\mathbf{B}$ by changing $P$ for two fixed knot types (set by $\leftrightarrow Q$). 
   Varying the helicity corresponds to varying amounts of winding of the lines of $\mathbf{B}$. 
   ({\bf a},{\bf b},{\bf c},{\bf d}): Trefoil knots with $Q_{\rm trefoil}$, ({\bf e},{\bf f},{\bf g},{\bf h}): Figure-8 knots with $Q_{\operatorname{fig-8}}$, with $Q$ functions as defined in Fig.~\ref{rational_map}.
 }
\label{helicity_variation}
\end{figure*}
 
{\it Tuning the helicity of a knotted field.}
The helicity of $\mathbf{B}$ can be tuned without changing the underlying knotted structure encoded in $\mathbf{B}$, as for rational maps \cite{sutcliffe_knots_2007}. 
The flow lines of $\mathbf{B}$ contained in the knotted tubes of constant $\chi$ in the neighborhood of $Q=0\iff\chi=1$, encode knots of the same type as the knot with $Q=0$.
However the degree of winding of these lines---and hence the helicity of $\mathbf{B}$---can be controlled by changing $P(u,v)$, as illustrated in Fig.~\ref{helicity_variation}.

Knotted fields encoding torus knots and links can be constructed from maps $\psi=P(u,v)/Q(u,v)$ with $P(u,v)=u^\alpha\,v^\beta$, $Q(u,v)=u^q+v^p$. The helicity of these fields can be varied without changing the underlying knotted structure by changing $\alpha,\beta$ in $P(u,v)=u^\alpha\,v^\beta$. 
The helicity of $\mathbf{B}$, being equal to the Hopf invariant \cite{whitehead_expression_1947,sutcliffe_knots_2007} of the map $\psi$, is $\mathcal{H}=\alpha\,p +\beta\,q$. The lines of the field $\mathbf{B}$ wind more for higher values of $\alpha,\beta$ as indicated by the higher values of helicity.  

Knotted fields encoding other knot types such as the figure-8 knot (Fig.~\ref{fig:psi_fibres}(d)) can be constructed from maps $\psi=u^\alpha/Q(u,u^\ast, v)$, and their helicity can be tuned by changing $\alpha$. 
Their helicity is $\mathcal{H}=\alpha \, \mathrm{deg}_v(Q)$ where $\mathrm{deg}_v(Q)$ is the highest power of $v$ appearing in $Q(u,u^\ast,v)$, and depends linearly on $\alpha$.
The helicity of these knotted fieds can be tuned further to yield negative values by substituting $P$ or $Q$ with their complex conjugates.

{\it Helicity of knotted flux tubes.}
Knotted flux tubes that are suitable initial conditions for isolated magnetic flux tubes in plasmas or vortex tubes in fluids, can be generated by restricting 
the knotted field $\mathbf{B}$ to the interior of a knotted tube: $\chi > \chi_{0}$ (see Fig.~\ref{fig:seifert_flux}). Such a knotted flux tube contains flux
$(1-\chi_{0})$, and its helicity can be calculated as in \cite{berger_topological_1984, chui_energy_1995} to be $\mathcal{H}_{\chi_0}= (1-\chi_0)^2\,\mathcal{H}_{\rm total}$. 
This is exactly the helicity of a flux tube \cite{berger_topological_1984, chui_energy_1995} for a uniformly twisted field with twist equal to $\mathcal{H}_{\rm total}$.

{\it Knotted fields with vanishing helicity.}
This construction can also generate vector fields which despite being knotted, have vanishing helicity.
If $\mathbf{B}$ has an underlying rational map with $\psi=P$ as the knotted complex scalar field (i.e.~$Q=1$), the vector potential $\mathbf{A}$ in Eq.~(\ref{A_nat}) is singularity-free, implying $\mathcal{H} = 0$.
This may appear surprising, but can be explained by the fact that the lines of $\mathbf{B}$ are of different handedness on different nested tori, such that that the total linking between all the lines (which is what helicity measures) cancels exactly. 
However, the lines of $\mathbf{B}$ in the interior of a given torus can have a non-vanishing helicity which is difficult to compute.

An alternative way to see the vanishing of helicity is to note that the helicity of the knotted field $\mathbf{B}$ must equal the Hopf invariant \cite{whitehead_expression_1947,sutcliffe_knots_2007} of the map given by $\psi$. 
In this case ($\psi=P(u,u^\ast,v,v^\ast)$), the set of $(u,u^\ast,v,v^\ast)$ such that $\psi = \infty$ is a null set, so the helicity of $\mathbf{B}$ vanishes.

{\it Summary.}
We have presented a general method for constructing physically viable knotted vector fields, encoding an arbitrary combination of knots woven together, and shown how to explicitly compute their helicity. 
Furthermore, we have shown how our construction can be used to obtain knotted flux tubes, and calculated their helicity.

Similar knotted solutions to Maxwell's equations \cite{kedia_tying_2013} have found application in the construction of topological solitons in magnetohydrodynamics \cite{thompson_constructing_2014} and simulations of resistive MHD flows \cite{smiet_self-organizing_2015}. 
The knotted vector fields presented here encode a much larger variety of knots, possess richer structure. 
These knotted fields could lead to novel topological solitons, and new insights about the role of helicity in the evolution of fluids and plasmas. 

Finally, our systematic procedure for calculating the helicity of the knotted field $\mathbf{B}$, may help accurately determine the Hopf charge of arbitrarily knotted Skyrme-Faddeev solitons\cite{battye_solitons_1999,sutcliffe_knots_2007} and help tighten the lower bound on how their minimum energy grows with their Hopf charge \cite{ward_hopf_1999}.\\

\begin{acknowledgments}
We are grateful to Ben Bode for a useful discussions. 
The authors are grateful to the KITP and the Newton Institute for hospitality during the early part of this work.  
W.T.M.I. acknowledges support from the National Science Foundation (NSF) and the Packard Foundation, D.F. and M.R.D. acknowledge support from the Leverhulme Trust Programme Grant `Scientific Properties of Complex Knots', and M.R.D. acknowledges support from the Royal Society, research was funded by a Leverhulme Trust Research Programme Grant. 
\end{acknowledgments}

\bibliography{initial_state}

\begin{thebibliography}{70}%
\makeatletter
\providecommand \@ifxundefined [1]{%
 \@ifx{#1\undefined}
}%
\providecommand \@ifnum [1]{%
 \ifnum #1\expandafter \@firstoftwo
 \else \expandafter \@secondoftwo
 \fi
}%
\providecommand \@ifx [1]{%
 \ifx #1\expandafter \@firstoftwo
 \else \expandafter \@secondoftwo
 \fi
}%
\providecommand \natexlab [1]{#1}%
\providecommand \enquote  [1]{``#1''}%
\providecommand \bibnamefont  [1]{#1}%
\providecommand \bibfnamefont [1]{#1}%
\providecommand \citenamefont [1]{#1}%
\providecommand \href@noop [0]{\@secondoftwo}%
\providecommand \href [0]{\begingroup \@sanitize@url \@href}%
\providecommand \@href[1]{\@@startlink{#1}\@@href}%
\providecommand \@@href[1]{\endgroup#1\@@endlink}%
\providecommand \@sanitize@url [0]{\catcode `\\12\catcode `\$12\catcode
  `\&12\catcode `\#12\catcode `\^12\catcode `\_12\catcode `\%12\relax}%
\providecommand \@@startlink[1]{}%
\providecommand \@@endlink[0]{}%
\providecommand \url  [0]{\begingroup\@sanitize@url \@url }%
\providecommand \@url [1]{\endgroup\@href {#1}{\urlprefix }}%
\providecommand \urlprefix  [0]{URL }%
\providecommand \Eprint [0]{\href }%
\providecommand \doibase [0]{http://dx.doi.org/}%
\providecommand \selectlanguage [0]{\@gobble}%
\providecommand \bibinfo  [0]{\@secondoftwo}%
\providecommand \bibfield  [0]{\@secondoftwo}%
\providecommand \translation [1]{[#1]}%
\providecommand \BibitemOpen [0]{}%
\providecommand \bibitemStop [0]{}%
\providecommand \bibitemNoStop [0]{.\EOS\space}%
\providecommand \EOS [0]{\spacefactor3000\relax}%
\providecommand \BibitemShut  [1]{\csname bibitem#1\endcsname}%
\let\auto@bib@innerbib\@empty
%</preamble>
\bibitem [{\citenamefont {Moffatt}(1969)}]{moffatt_degree_1969}%
  \BibitemOpen
  \bibfield  {author} {\bibinfo {author} {\bibfnamefont {H.~K.}\ \bibnamefont
  {Moffatt}},\ }\href {\doibase 10.1017/S0022112069000991} {\bibfield
  {journal} {\bibinfo  {journal} {Journal of Fluid Mechanics}\ }\textbf
  {\bibinfo {volume} {35}},\ \bibinfo {pages} {117} (\bibinfo {year}
  {1969})}\BibitemShut {NoStop}%
\bibitem [{\citenamefont {Enciso}\ and\ \citenamefont
  {Peralta-Salas}(2012)}]{enciso_knots_2012}%
  \BibitemOpen
  \bibfield  {author} {\bibinfo {author} {\bibfnamefont {A.}~\bibnamefont
  {Enciso}}\ and\ \bibinfo {author} {\bibfnamefont {D.}~\bibnamefont
  {Peralta-Salas}},\ }\href {\doibase 10.4007/annals.2012.175.1.9} {\bibfield
  {journal} {\bibinfo  {journal} {Annals of Mathematics}\ }\textbf {\bibinfo
  {volume} {175}},\ \bibinfo {pages} {345} (\bibinfo {year}
  {2012})}\BibitemShut {NoStop}%
\bibitem [{\citenamefont {Kleckner}\ and\ \citenamefont
  {Irvine}(2013)}]{kleckner_creation_2013}%
  \BibitemOpen
  \bibfield  {author} {\bibinfo {author} {\bibfnamefont {D.}~\bibnamefont
  {Kleckner}}\ and\ \bibinfo {author} {\bibfnamefont {W.~T.~M.}\ \bibnamefont
  {Irvine}},\ }\href {\doibase 10.1038/nphys2560} {\bibfield  {journal}
  {\bibinfo  {journal} {Nat Phys}\ }\textbf {\bibinfo {volume} {9}},\ \bibinfo
  {pages} {253} (\bibinfo {year} {2013})}\BibitemShut {NoStop}%
\bibitem [{\citenamefont {Scheeler}\ \emph {et~al.}(2014)\citenamefont
  {Scheeler}, \citenamefont {Kleckner}, \citenamefont {Proment}, \citenamefont
  {Kindlmann},\ and\ \citenamefont {Irvine}}]{scheeler_helicity_2014}%
  \BibitemOpen
  \bibfield  {author} {\bibinfo {author} {\bibfnamefont {M.~W.}\ \bibnamefont
  {Scheeler}}, \bibinfo {author} {\bibfnamefont {D.}~\bibnamefont {Kleckner}},
  \bibinfo {author} {\bibfnamefont {D.}~\bibnamefont {Proment}}, \bibinfo
  {author} {\bibfnamefont {G.~L.}\ \bibnamefont {Kindlmann}}, \ and\ \bibinfo
  {author} {\bibfnamefont {W.~T.~M.}\ \bibnamefont {Irvine}},\ }\href {\doibase
  10.1073/pnas.1407232111} {\bibfield  {journal} {\bibinfo  {journal} {PNAS}\
  }\textbf {\bibinfo {volume} {111}},\ \bibinfo {pages} {15350} (\bibinfo
  {year} {2014})}\BibitemShut {NoStop}%
\bibitem [{\citenamefont {Barenghi}(2007)}]{barenghi_knots_2007}%
  \BibitemOpen
  \bibfield  {author} {\bibinfo {author} {\bibfnamefont {C.~F.}\ \bibnamefont
  {Barenghi}},\ }\href {\doibase 10.1007/s00032-007-0069-5} {\bibfield
  {journal} {\bibinfo  {journal} {Milan J. Math.}\ }\textbf {\bibinfo {volume}
  {75}},\ \bibinfo {pages} {177} (\bibinfo {year} {2007})}\BibitemShut
  {NoStop}%
\bibitem [{\citenamefont {Kawaguchi}\ \emph {et~al.}(2008)\citenamefont
  {Kawaguchi}, \citenamefont {Nitta},\ and\ \citenamefont
  {Ueda}}]{kawaguchi_knots_2008}%
  \BibitemOpen
  \bibfield  {author} {\bibinfo {author} {\bibfnamefont {Y.}~\bibnamefont
  {Kawaguchi}}, \bibinfo {author} {\bibfnamefont {M.}~\bibnamefont {Nitta}}, \
  and\ \bibinfo {author} {\bibfnamefont {M.}~\bibnamefont {Ueda}},\ }\href
  {\doibase 10.1103/PhysRevLett.100.180403} {\bibfield  {journal} {\bibinfo
  {journal} {Phys. Rev. Lett.}\ }\textbf {\bibinfo {volume} {100}},\ \bibinfo
  {pages} {180403} (\bibinfo {year} {2008})}\BibitemShut {NoStop}%
\bibitem [{\citenamefont {Hall}\ \emph {et~al.}(2016)\citenamefont {Hall},
  \citenamefont {Ray}, \citenamefont {Tiurev}, \citenamefont {Ruokokoski},
  \citenamefont {Gheorghe},\ and\ \citenamefont
  {Möttönen}}]{hall_tying_2016}%
  \BibitemOpen
  \bibfield  {author} {\bibinfo {author} {\bibfnamefont {D.~S.}\ \bibnamefont
  {Hall}}, \bibinfo {author} {\bibfnamefont {M.~W.}\ \bibnamefont {Ray}},
  \bibinfo {author} {\bibfnamefont {K.}~\bibnamefont {Tiurev}}, \bibinfo
  {author} {\bibfnamefont {E.}~\bibnamefont {Ruokokoski}}, \bibinfo {author}
  {\bibfnamefont {A.~H.}\ \bibnamefont {Gheorghe}}, \ and\ \bibinfo {author}
  {\bibfnamefont {M.}~\bibnamefont {Möttönen}},\ }\href {\doibase
  10.1038/nphys3624} {\bibfield  {journal} {\bibinfo  {journal} {Nat Phys}\
  }\textbf {\bibinfo {volume} {12}},\ \bibinfo {pages} {478} (\bibinfo {year}
  {2016})}\BibitemShut {NoStop}%
\bibitem [{\citenamefont {Kleckner}\ \emph {et~al.}(2016)\citenamefont
  {Kleckner}, \citenamefont {Kauffman},\ and\ \citenamefont
  {Irvine}}]{kleckner_how_2016}%
  \BibitemOpen
  \bibfield  {author} {\bibinfo {author} {\bibfnamefont {D.}~\bibnamefont
  {Kleckner}}, \bibinfo {author} {\bibfnamefont {L.~H.}\ \bibnamefont
  {Kauffman}}, \ and\ \bibinfo {author} {\bibfnamefont {W.~T.~M.}\ \bibnamefont
  {Irvine}},\ }\href {\doibase 10.1038/nphys3679} {\bibfield  {journal}
  {\bibinfo  {journal} {Nat. Phys.}\ } (\bibinfo {year} {2016}),\
  10.1038/nphys3679}\BibitemShut {NoStop}%
\bibitem [{\citenamefont {Kamchatnov}(1982)}]{kamchatnov_topological_1982}%
  \BibitemOpen
  \bibfield  {author} {\bibinfo {author} {\bibfnamefont {A.~M.}\ \bibnamefont
  {Kamchatnov}},\ }\href@noop {} {\bibfield  {journal} {\bibinfo  {journal}
  {JETP}\ ,\ \bibinfo {pages} {117}} (\bibinfo {year} {1982})}\BibitemShut
  {NoStop}%
\bibitem [{\citenamefont {Moffatt}(1985)}]{moffatt_magnetostatic_1985}%
  \BibitemOpen
  \bibfield  {author} {\bibinfo {author} {\bibfnamefont {H.~K.}\ \bibnamefont
  {Moffatt}},\ }\href {\doibase 10.1017/S0022112085003251} {\bibfield
  {journal} {\bibinfo  {journal} {Journal of Fluid Mechanics}\ }\textbf
  {\bibinfo {volume} {159}},\ \bibinfo {pages} {359} (\bibinfo {year}
  {1985})}\BibitemShut {NoStop}%
\bibitem [{\citenamefont {Ranada}(1989)}]{ranada_topological_1989}%
  \BibitemOpen
  \bibfield  {author} {\bibinfo {author} {\bibfnamefont {A.~F.}\ \bibnamefont
  {Ranada}},\ }\href {\doibase 10.1007/BF00401864} {\bibfield  {journal}
  {\bibinfo  {journal} {Lett. Math. Phys.}\ }\textbf {\bibinfo {volume} {18}},\
  \bibinfo {pages} {97} (\bibinfo {year} {1989})}\BibitemShut {NoStop}%
\bibitem [{\citenamefont {Chui}\ and\ \citenamefont
  {Moffatt}(1995)}]{chui_energy_1995}%
  \BibitemOpen
  \bibfield  {author} {\bibinfo {author} {\bibfnamefont {A.~Y.~K.}\
  \bibnamefont {Chui}}\ and\ \bibinfo {author} {\bibfnamefont {H.~K.}\
  \bibnamefont {Moffatt}},\ }\href {\doibase 10.1098/rspa.1995.0146} {\bibfield
   {journal} {\bibinfo  {journal} {Proc. R. Soc. A}\ }\textbf {\bibinfo
  {volume} {451}},\ \bibinfo {pages} {609} (\bibinfo {year}
  {1995})}\BibitemShut {NoStop}%
\bibitem [{\citenamefont {Irvine}\ and\ \citenamefont
  {Bouwmeester}(2008)}]{irvine_linked_2008}%
  \BibitemOpen
  \bibfield  {author} {\bibinfo {author} {\bibfnamefont {W.~T.~M.}\
  \bibnamefont {Irvine}}\ and\ \bibinfo {author} {\bibfnamefont
  {D.}~\bibnamefont {Bouwmeester}},\ }\href {\doibase 10.1038/nphys1056}
  {\bibfield  {journal} {\bibinfo  {journal} {Nat. Phys.}\ }\textbf {\bibinfo
  {volume} {4}},\ \bibinfo {pages} {716} (\bibinfo {year} {2008})}\BibitemShut
  {NoStop}%
\bibitem [{\citenamefont {Irvine}(2010)}]{irvine_linked_2010}%
  \BibitemOpen
  \bibfield  {author} {\bibinfo {author} {\bibfnamefont {W.~T.~M.}\
  \bibnamefont {Irvine}},\ }\href {\doibase 10.1088/1751-8113/43/38/385203}
  {\bibfield  {journal} {\bibinfo  {journal} {J. Phys. A: Math. Theor.}\
  }\textbf {\bibinfo {volume} {43}},\ \bibinfo {pages} {385203} (\bibinfo
  {year} {2010})}\BibitemShut {NoStop}%
\bibitem [{\citenamefont {Arrayas}\ and\ \citenamefont
  {Trueba}(2015)}]{arrayas_class_2015}%
  \BibitemOpen
  \bibfield  {author} {\bibinfo {author} {\bibfnamefont {M.}~\bibnamefont
  {Arrayas}}\ and\ \bibinfo {author} {\bibfnamefont {J.~L.}\ \bibnamefont
  {Trueba}},\ }\href {\doibase 10.1088/1751-8113/48/2/025203} {\bibfield
  {journal} {\bibinfo  {journal} {J. Phys. A: Math. Theor.}\ }\textbf {\bibinfo
  {volume} {48}},\ \bibinfo {pages} {025203} (\bibinfo {year}
  {2015})}\BibitemShut {NoStop}%
\bibitem [{\citenamefont {Thompson}\ \emph {et~al.}(2014)\citenamefont
  {Thompson}, \citenamefont {Swearngin}, \citenamefont {Wickes},\ and\
  \citenamefont {Bouwmeester}}]{thompson_constructing_2014}%
  \BibitemOpen
  \bibfield  {author} {\bibinfo {author} {\bibfnamefont {A.}~\bibnamefont
  {Thompson}}, \bibinfo {author} {\bibfnamefont {J.}~\bibnamefont {Swearngin}},
  \bibinfo {author} {\bibfnamefont {A.}~\bibnamefont {Wickes}}, \ and\ \bibinfo
  {author} {\bibfnamefont {D.}~\bibnamefont {Bouwmeester}},\ }\href {\doibase
  10.1103/PhysRevE.89.043104} {\bibfield  {journal} {\bibinfo  {journal} {Phys.
  Rev. E}\ }\textbf {\bibinfo {volume} {89}},\ \bibinfo {pages} {043104}
  (\bibinfo {year} {2014})}\BibitemShut {NoStop}%
\bibitem [{\citenamefont {Smiet}\ \emph {et~al.}(2015)\citenamefont {Smiet},
  \citenamefont {Candelaresi}, \citenamefont {Thompson}, \citenamefont
  {Swearngin}, \citenamefont {Dalhuisen},\ and\ \citenamefont
  {Bouwmeester}}]{smiet_self-organizing_2015}%
  \BibitemOpen
  \bibfield  {author} {\bibinfo {author} {\bibfnamefont {C.~B.}\ \bibnamefont
  {Smiet}}, \bibinfo {author} {\bibfnamefont {S.}~\bibnamefont {Candelaresi}},
  \bibinfo {author} {\bibfnamefont {A.}~\bibnamefont {Thompson}}, \bibinfo
  {author} {\bibfnamefont {J.}~\bibnamefont {Swearngin}}, \bibinfo {author}
  {\bibfnamefont {J.~W.}\ \bibnamefont {Dalhuisen}}, \ and\ \bibinfo {author}
  {\bibfnamefont {D.}~\bibnamefont {Bouwmeester}},\ }\href {\doibase
  10.1103/PhysRevLett.115.095001} {\bibfield  {journal} {\bibinfo  {journal}
  {Phys. Rev. Lett.}\ }\textbf {\bibinfo {volume} {115}},\ \bibinfo {pages}
  {095001} (\bibinfo {year} {2015})}\BibitemShut {NoStop}%
\bibitem [{\citenamefont {Ranada}\ \emph {et~al.}(2000)\citenamefont {Ranada},
  \citenamefont {Soler},\ and\ \citenamefont {Trueba}}]{ranada_ball_2000}%
  \BibitemOpen
  \bibfield  {author} {\bibinfo {author} {\bibfnamefont {A.~F.}\ \bibnamefont
  {Ranada}}, \bibinfo {author} {\bibfnamefont {M.}~\bibnamefont {Soler}}, \
  and\ \bibinfo {author} {\bibfnamefont {J.~L.}\ \bibnamefont {Trueba}},\
  }\href {\doibase 10.1103/PhysRevE.62.7181} {\bibfield  {journal} {\bibinfo
  {journal} {Phys. Rev. E}\ }\textbf {\bibinfo {volume} {62}},\ \bibinfo
  {pages} {7181} (\bibinfo {year} {2000})}\BibitemShut {NoStop}%
\bibitem [{\citenamefont {Kedia}\ \emph {et~al.}(2013)\citenamefont {Kedia},
  \citenamefont {Bialynicki-Birula}, \citenamefont {Peralta-Salas},\ and\
  \citenamefont {Irvine}}]{kedia_tying_2013}%
  \BibitemOpen
  \bibfield  {author} {\bibinfo {author} {\bibfnamefont {H.}~\bibnamefont
  {Kedia}}, \bibinfo {author} {\bibfnamefont {I.}~\bibnamefont
  {Bialynicki-Birula}}, \bibinfo {author} {\bibfnamefont {D.}~\bibnamefont
  {Peralta-Salas}}, \ and\ \bibinfo {author} {\bibfnamefont {W.~T.~M.}\
  \bibnamefont {Irvine}},\ }\href {\doibase 10.1103/PhysRevLett.111.150404}
  {\bibfield  {journal} {\bibinfo  {journal} {Phys. Rev. Lett.}\ }\textbf
  {\bibinfo {volume} {111}},\ \bibinfo {pages} {150404} (\bibinfo {year}
  {2013})}\BibitemShut {NoStop}%
\bibitem [{\citenamefont {Tkalec}\ \emph {et~al.}(2011)\citenamefont {Tkalec},
  \citenamefont {Ravnik}, \citenamefont {Čopar}, \citenamefont {Žumer},\ and\
  \citenamefont {Muševič}}]{tkalec_reconfigurable_2011}%
  \BibitemOpen
  \bibfield  {author} {\bibinfo {author} {\bibfnamefont {U.}~\bibnamefont
  {Tkalec}}, \bibinfo {author} {\bibfnamefont {M.}~\bibnamefont {Ravnik}},
  \bibinfo {author} {\bibfnamefont {S.}~\bibnamefont {Čopar}}, \bibinfo
  {author} {\bibfnamefont {S.}~\bibnamefont {Žumer}}, \ and\ \bibinfo {author}
  {\bibfnamefont {I.}~\bibnamefont {Muševič}},\ }\href {\doibase
  10.1126/science.1205705} {\bibfield  {journal} {\bibinfo  {journal}
  {Science}\ }\textbf {\bibinfo {volume} {333}},\ \bibinfo {pages} {62}
  (\bibinfo {year} {2011})}\BibitemShut {NoStop}%
\bibitem [{\citenamefont {Alexander}\ \emph {et~al.}(2012)\citenamefont
  {Alexander}, \citenamefont {Chen}, \citenamefont {Matsumoto},\ and\
  \citenamefont {Kamien}}]{alexander_colloquium:_2012}%
  \BibitemOpen
  \bibfield  {author} {\bibinfo {author} {\bibfnamefont {G.~P.}\ \bibnamefont
  {Alexander}}, \bibinfo {author} {\bibfnamefont {B.~G.-g.}\ \bibnamefont
  {Chen}}, \bibinfo {author} {\bibfnamefont {E.~A.}\ \bibnamefont {Matsumoto}},
  \ and\ \bibinfo {author} {\bibfnamefont {R.~D.}\ \bibnamefont {Kamien}},\
  }\href {\doibase 10.1103/RevModPhys.84.497} {\bibfield  {journal} {\bibinfo
  {journal} {Rev. Mod. Phys.}\ }\textbf {\bibinfo {volume} {84}},\ \bibinfo
  {pages} {497} (\bibinfo {year} {2012})}\BibitemShut {NoStop}%
\bibitem [{\citenamefont {Machon}\ and\ \citenamefont
  {Alexander}(2013)}]{machon_knots_2013}%
  \BibitemOpen
  \bibfield  {author} {\bibinfo {author} {\bibfnamefont {T.}~\bibnamefont
  {Machon}}\ and\ \bibinfo {author} {\bibfnamefont {G.~P.}\ \bibnamefont
  {Alexander}},\ }\href {\doibase 10.1073/pnas.1308225110} {\bibfield
  {journal} {\bibinfo  {journal} {PNAS}\ }\textbf {\bibinfo {volume} {110}},\
  \bibinfo {pages} {14174} (\bibinfo {year} {2013})}\BibitemShut {NoStop}%
\bibitem [{\citenamefont {Martinez}\ \emph {et~al.}(2014)\citenamefont
  {Martinez}, \citenamefont {Ravnik}, \citenamefont {Lucero}, \citenamefont
  {Visvanathan}, \citenamefont {Zumer},\ and\ \citenamefont
  {Smalyukh}}]{martinez_mutually_2014}%
  \BibitemOpen
  \bibfield  {author} {\bibinfo {author} {\bibfnamefont {A.}~\bibnamefont
  {Martinez}}, \bibinfo {author} {\bibfnamefont {M.}~\bibnamefont {Ravnik}},
  \bibinfo {author} {\bibfnamefont {B.}~\bibnamefont {Lucero}}, \bibinfo
  {author} {\bibfnamefont {R.}~\bibnamefont {Visvanathan}}, \bibinfo {author}
  {\bibfnamefont {S.}~\bibnamefont {Zumer}}, \ and\ \bibinfo {author}
  {\bibfnamefont {I.~I.}\ \bibnamefont {Smalyukh}},\ }\href {\doibase
  10.1038/nmat3840} {\bibfield  {journal} {\bibinfo  {journal} {Nat. Mat.}\
  }\textbf {\bibinfo {volume} {13}},\ \bibinfo {pages} {258} (\bibinfo {year}
  {2014})}\BibitemShut {NoStop}%
\bibitem [{\citenamefont {Berry}\ and\ \citenamefont
  {Dennis}(2000)}]{berry_phase_2000}%
  \BibitemOpen
  \bibfield  {author} {\bibinfo {author} {\bibfnamefont {M.~V.}\ \bibnamefont
  {Berry}}\ and\ \bibinfo {author} {\bibfnamefont {M.~R.}\ \bibnamefont
  {Dennis}},\ }\href {\doibase 10.1098/rspa.2000.0602} {\bibfield  {journal}
  {\bibinfo  {journal} {Proc. R. Soc. A}\ }\textbf {\bibinfo {volume} {456}},\
  \bibinfo {pages} {2059} (\bibinfo {year} {2000})}\BibitemShut {NoStop}%
\bibitem [{\citenamefont {Dennis}\ \emph {et~al.}(2010)\citenamefont {Dennis},
  \citenamefont {King}, \citenamefont {Jack}, \citenamefont {O’Holleran},\
  and\ \citenamefont {Padgett}}]{dennis_isolated_2010}%
  \BibitemOpen
  \bibfield  {author} {\bibinfo {author} {\bibfnamefont {M.~R.}\ \bibnamefont
  {Dennis}}, \bibinfo {author} {\bibfnamefont {R.~P.}\ \bibnamefont {King}},
  \bibinfo {author} {\bibfnamefont {B.}~\bibnamefont {Jack}}, \bibinfo {author}
  {\bibfnamefont {K.}~\bibnamefont {O’Holleran}}, \ and\ \bibinfo {author}
  {\bibfnamefont {M.~J.}\ \bibnamefont {Padgett}},\ }\href {\doibase
  10.1038/nphys1504} {\bibfield  {journal} {\bibinfo  {journal} {Nat. Phys.}\
  }\textbf {\bibinfo {volume} {6}},\ \bibinfo {pages} {118} (\bibinfo {year}
  {2010})}\BibitemShut {NoStop}%
\bibitem [{\citenamefont {Faddeev}\ and\ \citenamefont
  {Niemi}(1997)}]{faddeev_stable_1997}%
  \BibitemOpen
  \bibfield  {author} {\bibinfo {author} {\bibfnamefont {L.}~\bibnamefont
  {Faddeev}}\ and\ \bibinfo {author} {\bibfnamefont {A.~J.}\ \bibnamefont
  {Niemi}},\ }\href {\doibase 10.1038/387058a0} {\bibfield  {journal} {\bibinfo
   {journal} {Nature}\ }\textbf {\bibinfo {volume} {387}},\ \bibinfo {pages}
  {58} (\bibinfo {year} {1997})}\BibitemShut {NoStop}%
\bibitem [{\citenamefont {Battye}\ and\ \citenamefont
  {Sutcliffe}(1999)}]{battye_solitons_1999}%
  \BibitemOpen
  \bibfield  {author} {\bibinfo {author} {\bibfnamefont {R.~A.}\ \bibnamefont
  {Battye}}\ and\ \bibinfo {author} {\bibfnamefont {P.~M.}\ \bibnamefont
  {Sutcliffe}},\ }\href {\doibase 10.1098/rspa.1999.0502} {\bibfield  {journal}
  {\bibinfo  {journal} {Proc. R. Soc. A}\ }\textbf {\bibinfo {volume} {455}},\
  \bibinfo {pages} {4305} (\bibinfo {year} {1999})}\BibitemShut {NoStop}%
\bibitem [{\citenamefont {Babaev}\ \emph {et~al.}(2002)\citenamefont {Babaev},
  \citenamefont {Faddeev},\ and\ \citenamefont {Niemi}}]{babaev_hidden_2002}%
  \BibitemOpen
  \bibfield  {author} {\bibinfo {author} {\bibfnamefont {E.}~\bibnamefont
  {Babaev}}, \bibinfo {author} {\bibfnamefont {L.~D.}\ \bibnamefont {Faddeev}},
  \ and\ \bibinfo {author} {\bibfnamefont {A.~J.}\ \bibnamefont {Niemi}},\
  }\href {\doibase 10.1103/PhysRevB.65.100512} {\bibfield  {journal} {\bibinfo
  {journal} {Phys. Rev. B}\ }\textbf {\bibinfo {volume} {65}},\ \bibinfo
  {pages} {100512} (\bibinfo {year} {2002})}\BibitemShut {NoStop}%
\bibitem [{\citenamefont {Sutcliffe}(2007)}]{sutcliffe_knots_2007}%
  \BibitemOpen
  \bibfield  {author} {\bibinfo {author} {\bibfnamefont {P.}~\bibnamefont
  {Sutcliffe}},\ }\href {\doibase 10.1098/rspa.2007.0038} {\bibfield  {journal}
  {\bibinfo  {journal} {Proc. R. Soc. A}\ }\textbf {\bibinfo {volume} {463}},\
  \bibinfo {pages} {3001} (\bibinfo {year} {2007})}\BibitemShut {NoStop}%
\bibitem [{\citenamefont {Babaev}(2002)}]{babaev_dual_2002}%
  \BibitemOpen
  \bibfield  {author} {\bibinfo {author} {\bibfnamefont {E.}~\bibnamefont
  {Babaev}},\ }\href {\doibase 10.1103/PhysRevLett.88.177002} {\bibfield
  {journal} {\bibinfo  {journal} {Phys. Rev. Lett.}\ }\textbf {\bibinfo
  {volume} {88}},\ \bibinfo {pages} {177002} (\bibinfo {year}
  {2002})}\BibitemShut {NoStop}%
\bibitem [{\citenamefont {Babaev}(2009)}]{babaev_non-meissner_2009}%
  \BibitemOpen
  \bibfield  {author} {\bibinfo {author} {\bibfnamefont {E.}~\bibnamefont
  {Babaev}},\ }\href {\doibase 10.1103/PhysRevB.79.104506} {\bibfield
  {journal} {\bibinfo  {journal} {Phys. Rev. B}\ }\textbf {\bibinfo {volume}
  {79}},\ \bibinfo {pages} {104506} (\bibinfo {year} {2009})}\BibitemShut
  {NoStop}%
\bibitem [{\citenamefont {Helmholtz}(1858)}]{helmholtz_uber_1858}%
  \BibitemOpen
  \bibfield  {author} {\bibinfo {author} {\bibfnamefont {H.}~\bibnamefont
  {Helmholtz}},\ }\href {https://eudml.org/doc/147720} {\bibfield  {journal}
  {\bibinfo  {journal} {J. Reine Angew. Math.}\ }\textbf {\bibinfo {volume}
  {55}},\ \bibinfo {pages} {25} (\bibinfo {year} {1858})}\BibitemShut {NoStop}%
\bibitem [{\citenamefont {Thomson}(1869)}]{thomson_vortex_1869}%
  \BibitemOpen
  \bibfield  {author} {\bibinfo {author} {\bibfnamefont {W.}~\bibnamefont
  {Thomson}},\ }\href {\doibase 10.1017/S0370164600045430} {\bibfield
  {journal} {\bibinfo  {journal} {Proc. R. Soc. Edinburgh}\ }\textbf {\bibinfo
  {volume} {6}},\ \bibinfo {pages} {94} (\bibinfo {year} {1869})}\BibitemShut
  {NoStop}%
\bibitem [{\citenamefont {Woltjer}(1958)}]{woltjer_theorem_1958}%
  \BibitemOpen
  \bibfield  {author} {\bibinfo {author} {\bibfnamefont {L.}~\bibnamefont
  {Woltjer}},\ }\href@noop {} {\bibfield  {journal} {\bibinfo  {journal}
  {PNAS}\ }\textbf {\bibinfo {volume} {44}},\ \bibinfo {pages} {489} (\bibinfo
  {year} {1958})}\BibitemShut {NoStop}%
\bibitem [{\citenamefont {Chandrasekhar}\ and\ \citenamefont
  {Woltjer}(1958)}]{chandrasekhar_force-free_1958}%
  \BibitemOpen
  \bibfield  {author} {\bibinfo {author} {\bibfnamefont {S.}~\bibnamefont
  {Chandrasekhar}}\ and\ \bibinfo {author} {\bibfnamefont {L.}~\bibnamefont
  {Woltjer}},\ }\href {http://www.osti.gov/scitech/biblio/4333960} {\bibfield
  {journal} {\bibinfo  {journal} {PNAS}\ }\textbf {\bibinfo {volume} {44}}
  (\bibinfo {year} {1958})}\BibitemShut {NoStop}%
\bibitem [{\citenamefont {Newcomb}(1958)}]{newcomb_motion_1958}%
  \BibitemOpen
  \bibfield  {author} {\bibinfo {author} {\bibfnamefont {W.~A.}\ \bibnamefont
  {Newcomb}},\ }\href {\doibase 10.1016/0003-4916(58)90024-1} {\bibfield
  {journal} {\bibinfo  {journal} {Annals of Physics}\ }\textbf {\bibinfo
  {volume} {3}},\ \bibinfo {pages} {347} (\bibinfo {year} {1958})}\BibitemShut
  {NoStop}%
\bibitem [{\citenamefont {Monchaux}\ \emph {et~al.}(2009)\citenamefont
  {Monchaux}, \citenamefont {Berhanu}, \citenamefont {Aumaître}, \citenamefont
  {Chiffaudel}, \citenamefont {Daviaud}, \citenamefont {Dubrulle},
  \citenamefont {Ravelet}, \citenamefont {Fauve}, \citenamefont {Mordant},
  \citenamefont {Pétrélis}, \citenamefont {Bourgoin}, \citenamefont {Odier},
  \citenamefont {Pinton}, \citenamefont {Plihon},\ and\ \citenamefont
  {Volk}}]{monchaux_von_2009}%
  \BibitemOpen
  \bibfield  {author} {\bibinfo {author} {\bibfnamefont {R.}~\bibnamefont
  {Monchaux}}, \bibinfo {author} {\bibfnamefont {M.}~\bibnamefont {Berhanu}},
  \bibinfo {author} {\bibfnamefont {S.}~\bibnamefont {Aumaître}}, \bibinfo
  {author} {\bibfnamefont {A.}~\bibnamefont {Chiffaudel}}, \bibinfo {author}
  {\bibfnamefont {F.}~\bibnamefont {Daviaud}}, \bibinfo {author} {\bibfnamefont
  {B.}~\bibnamefont {Dubrulle}}, \bibinfo {author} {\bibfnamefont
  {F.}~\bibnamefont {Ravelet}}, \bibinfo {author} {\bibfnamefont
  {S.}~\bibnamefont {Fauve}}, \bibinfo {author} {\bibfnamefont
  {N.}~\bibnamefont {Mordant}}, \bibinfo {author} {\bibfnamefont
  {F.}~\bibnamefont {Pétrélis}}, \bibinfo {author} {\bibfnamefont
  {M.}~\bibnamefont {Bourgoin}}, \bibinfo {author} {\bibfnamefont
  {P.}~\bibnamefont {Odier}}, \bibinfo {author} {\bibfnamefont {J.-F.}\
  \bibnamefont {Pinton}}, \bibinfo {author} {\bibfnamefont {N.}~\bibnamefont
  {Plihon}}, \ and\ \bibinfo {author} {\bibfnamefont {R.}~\bibnamefont
  {Volk}},\ }\href {\doibase 10.1063/1.3085724} {\bibfield  {journal} {\bibinfo
   {journal} {Phys. Fluids}\ }\textbf {\bibinfo {volume} {21}},\ \bibinfo
  {pages} {035108} (\bibinfo {year} {2009})}\BibitemShut {NoStop}%
\bibitem [{\citenamefont {Steenbeck}\ \emph {et~al.}(2014)\citenamefont
  {Steenbeck}, \citenamefont {Krause},\ and\ \citenamefont
  {Rädler}}]{steenbeck_berechnung_2014}%
  \BibitemOpen
  \bibfield  {author} {\bibinfo {author} {\bibfnamefont {M.}~\bibnamefont
  {Steenbeck}}, \bibinfo {author} {\bibfnamefont {F.}~\bibnamefont {Krause}}, \
  and\ \bibinfo {author} {\bibfnamefont {K.-H.}\ \bibnamefont {Rädler}},\
  }\href {\doibase 10.1515/zna-1966-0401} {\bibfield  {journal} {\bibinfo
  {journal} {Z. Naturforsch. A Phys. Sci.}\ }\textbf {\bibinfo {volume} {21}},\
  \bibinfo {pages} {369} (\bibinfo {year} {2014})}\BibitemShut {NoStop}%
\bibitem [{\citenamefont {Moffatt}(2014)}]{moffatt_helicity_2014}%
  \BibitemOpen
  \bibfield  {author} {\bibinfo {author} {\bibfnamefont {H.~K.}\ \bibnamefont
  {Moffatt}},\ }\href {\doibase 10.1073/pnas.1400277111} {\bibfield  {journal}
  {\bibinfo  {journal} {PNAS}\ }\textbf {\bibinfo {volume} {111}},\ \bibinfo
  {pages} {3663} (\bibinfo {year} {2014})}\BibitemShut {NoStop}%
\bibitem [{\citenamefont {Arnold}(1974)}]{arnold_asymptotic_1974}%
  \BibitemOpen
  \bibfield  {author} {\bibinfo {author} {\bibfnamefont {V.~I.}\ \bibnamefont
  {Arnold}},\ }in\ \href
  {http://link.springer.com/chapter/10.1007/978-3-642-31031-7_32} {\emph
  {\bibinfo {booktitle} {Vladimir {I}. {Arnold} - {Collected} {Works}}}},\
  \bibinfo {series and number} {\bibinfo {number} {2}}\ (\bibinfo  {publisher}
  {Springer Berlin Heidelberg},\ \bibinfo {year} {1974})\ pp.\ \bibinfo {pages}
  {357--375}\BibitemShut {NoStop}%
\bibitem [{\citenamefont {Freedman}(1988)}]{freedman_note_1988}%
  \BibitemOpen
  \bibfield  {author} {\bibinfo {author} {\bibfnamefont {M.~H.}\ \bibnamefont
  {Freedman}},\ }\href {\doibase 10.1017/S002211208800309X} {\bibfield
  {journal} {\bibinfo  {journal} {J. Fluid Mech.}\ }\textbf {\bibinfo {volume}
  {194}},\ \bibinfo {pages} {549} (\bibinfo {year} {1988})}\BibitemShut
  {NoStop}%
\bibitem [{\citenamefont {Rogers}\ and\ \citenamefont
  {Moin}(1987)}]{rogers_helicity_1987}%
  \BibitemOpen
  \bibfield  {author} {\bibinfo {author} {\bibfnamefont {M.~M.}\ \bibnamefont
  {Rogers}}\ and\ \bibinfo {author} {\bibfnamefont {P.}~\bibnamefont {Moin}},\
  }\href {\doibase 10.1063/1.866030} {\bibfield  {journal} {\bibinfo  {journal}
  {Physics of Fluids (1958-1988)}\ }\textbf {\bibinfo {volume} {30}},\ \bibinfo
  {pages} {2662} (\bibinfo {year} {1987})}\BibitemShut {NoStop}%
\bibitem [{\citenamefont {Wallace}\ \emph {et~al.}(1992)\citenamefont
  {Wallace}, \citenamefont {Balint},\ and\ \citenamefont
  {Ong}}]{wallace_experimental_1992}%
  \BibitemOpen
  \bibfield  {author} {\bibinfo {author} {\bibfnamefont {J.~M.}\ \bibnamefont
  {Wallace}}, \bibinfo {author} {\bibfnamefont {J.-L.}\ \bibnamefont {Balint}},
  \ and\ \bibinfo {author} {\bibfnamefont {L.}~\bibnamefont {Ong}},\ }\href
  {\doibase 10.1063/1.858371} {\bibfield  {journal} {\bibinfo  {journal}
  {Physics of Fluids A: Fluid Dynamics (1989-1993)}\ }\textbf {\bibinfo
  {volume} {4}},\ \bibinfo {pages} {2013} (\bibinfo {year} {1992})}\BibitemShut
  {NoStop}%
\bibitem [{\citenamefont {Moffatt}(1990)}]{moffatt_energy_1990}%
  \BibitemOpen
  \bibfield  {author} {\bibinfo {author} {\bibfnamefont {H.~K.}\ \bibnamefont
  {Moffatt}},\ }\href {\doibase 10.1038/347367a0} {\bibfield  {journal}
  {\bibinfo  {journal} {Nature}\ }\textbf {\bibinfo {volume} {347}},\ \bibinfo
  {pages} {367} (\bibinfo {year} {1990})}\BibitemShut {NoStop}%
\bibitem [{\citenamefont {Katritch}\ \emph {et~al.}(1996)\citenamefont
  {Katritch}, \citenamefont {Bednar}, \citenamefont {Michoud}, \citenamefont
  {Scharein}, \citenamefont {Dubochet},\ and\ \citenamefont
  {Stasiak}}]{katritch_geometry_1996}%
  \BibitemOpen
  \bibfield  {author} {\bibinfo {author} {\bibfnamefont {V.}~\bibnamefont
  {Katritch}}, \bibinfo {author} {\bibfnamefont {J.}~\bibnamefont {Bednar}},
  \bibinfo {author} {\bibfnamefont {D.}~\bibnamefont {Michoud}}, \bibinfo
  {author} {\bibfnamefont {R.~G.}\ \bibnamefont {Scharein}}, \bibinfo {author}
  {\bibfnamefont {J.}~\bibnamefont {Dubochet}}, \ and\ \bibinfo {author}
  {\bibfnamefont {A.}~\bibnamefont {Stasiak}},\ }\href {\doibase
  10.1038/384142a0} {\bibfield  {journal} {\bibinfo  {journal} {Nature}\
  }\textbf {\bibinfo {volume} {384}},\ \bibinfo {pages} {142} (\bibinfo {year}
  {1996})}\BibitemShut {NoStop}%
\bibitem [{\citenamefont {Pieranski}\ and\ \citenamefont
  {Przybyl}(2001)}]{pieranski_ideal_2001}%
  \BibitemOpen
  \bibfield  {author} {\bibinfo {author} {\bibfnamefont {P.}~\bibnamefont
  {Pieranski}}\ and\ \bibinfo {author} {\bibfnamefont {S.}~\bibnamefont
  {Przybyl}},\ }\href {\doibase 10.1103/PhysRevE.64.031801} {\bibfield
  {journal} {\bibinfo  {journal} {Phys. Rev. E}\ }\textbf {\bibinfo {volume}
  {64}},\ \bibinfo {pages} {031801} (\bibinfo {year} {2001})}\BibitemShut
  {NoStop}%
\bibitem [{\citenamefont {Bajer}\ \emph {et~al.}(2013)\citenamefont {Bajer},
  \citenamefont {Kimura},\ and\ \citenamefont {Moffatt}}]{bajer_preface_2013}%
  \BibitemOpen
  \bibfield  {author} {\bibinfo {author} {\bibfnamefont {K.}~\bibnamefont
  {Bajer}}, \bibinfo {author} {\bibfnamefont {Y.}~\bibnamefont {Kimura}}, \
  and\ \bibinfo {author} {\bibfnamefont {H.~K.}\ \bibnamefont {Moffatt}},\
  }\href {\doibase 10.1016/j.piutam.2013.03.001} {\bibfield  {journal}
  {\bibinfo  {journal} {Procedia IUTAM}\ }\textbf {\bibinfo {volume} {7}},\
  \bibinfo {pages} {1} (\bibinfo {year} {2013})}\BibitemShut {NoStop}%
\bibitem [{\citenamefont {Stasiak}\ \emph {et~al.}(2013)\citenamefont
  {Stasiak}, \citenamefont {Bates}, \citenamefont {Buck}, \citenamefont
  {Harris},\ and\ \citenamefont {Sumners}}]{stasiak_topological_2013}%
  \BibitemOpen
  \bibfield  {author} {\bibinfo {author} {\bibfnamefont {A.}~\bibnamefont
  {Stasiak}}, \bibinfo {author} {\bibfnamefont {A.~D.}\ \bibnamefont {Bates}},
  \bibinfo {author} {\bibfnamefont {D.~E.}\ \bibnamefont {Buck}}, \bibinfo
  {author} {\bibfnamefont {S.~A.}\ \bibnamefont {Harris}}, \ and\ \bibinfo
  {author} {\bibfnamefont {D.~W.}\ \bibnamefont {Sumners}},\ }\href {\doibase
  10.1042/BST20130006} {\bibfield  {journal} {\bibinfo  {journal} {Biochemical
  Society Transactions}\ }\textbf {\bibinfo {volume} {41}},\ \bibinfo {pages}
  {491} (\bibinfo {year} {2013})}\BibitemShut {NoStop}%
\bibitem [{\citenamefont {Buniy}\ and\ \citenamefont
  {Kephart}(2003)}]{buniy_model_2003}%
  \BibitemOpen
  \bibfield  {author} {\bibinfo {author} {\bibfnamefont {R.~V.}\ \bibnamefont
  {Buniy}}\ and\ \bibinfo {author} {\bibfnamefont {T.~W.}\ \bibnamefont
  {Kephart}},\ }\href {\doibase 10.1016/j.physletb.2003.09.081} {\bibfield
  {journal} {\bibinfo  {journal} {Physics Letters B}\ }\textbf {\bibinfo
  {volume} {576}},\ \bibinfo {pages} {127} (\bibinfo {year}
  {2003})}\BibitemShut {NoStop}%
\bibitem [{\citenamefont {Buniy}\ and\ \citenamefont
  {Kephart}(2005)}]{buniy_glueballs_2005}%
  \BibitemOpen
  \bibfield  {author} {\bibinfo {author} {\bibfnamefont {R.~V.}\ \bibnamefont
  {Buniy}}\ and\ \bibinfo {author} {\bibfnamefont {T.~W.}\ \bibnamefont
  {Kephart}},\ }\href {\doibase 10.1142/S0217751X05024146} {\bibfield
  {journal} {\bibinfo  {journal} {Int. J. Mod. Phys. A}\ }\textbf {\bibinfo
  {volume} {20}},\ \bibinfo {pages} {1252} (\bibinfo {year}
  {2005})}\BibitemShut {NoStop}%
\bibitem [{\citenamefont {Buniy}\ \emph {et~al.}(2014)\citenamefont {Buniy},
  \citenamefont {Cantarella}, \citenamefont {Kephart},\ and\ \citenamefont
  {Rawdon}}]{buniy_tight_2014}%
  \BibitemOpen
  \bibfield  {author} {\bibinfo {author} {\bibfnamefont {R.~V.}\ \bibnamefont
  {Buniy}}, \bibinfo {author} {\bibfnamefont {J.}~\bibnamefont {Cantarella}},
  \bibinfo {author} {\bibfnamefont {T.~W.}\ \bibnamefont {Kephart}}, \ and\
  \bibinfo {author} {\bibfnamefont {E.~J.}\ \bibnamefont {Rawdon}},\ }\href
  {\doibase 10.1103/PhysRevD.89.054513} {\bibfield  {journal} {\bibinfo
  {journal} {Phys. Rev. D}\ }\textbf {\bibinfo {volume} {89}},\ \bibinfo
  {pages} {054513} (\bibinfo {year} {2014})}\BibitemShut {NoStop}%
\bibitem [{\citenamefont {Milnor}(1969)}]{milnor_singular_1969}%
  \BibitemOpen
  \bibfield  {author} {\bibinfo {author} {\bibfnamefont {J.}~\bibnamefont
  {Milnor}},\ }\href {http://press.princeton.edu/titles/1570.html} {\emph
  {\bibinfo {title} {Singular {Points} of {Complex} {Hypersurfaces}.}}},\
  Annals of {Mathematics} {Studies}\ (\bibinfo  {publisher} {Princeton
  University Press},\ \bibinfo {year} {1969})\BibitemShut {NoStop}%
\bibitem [{\citenamefont {Brauner}(1928)}]{brauner_verhalten_1928}%
  \BibitemOpen
  \bibfield  {author} {\bibinfo {author} {\bibfnamefont {K.}~\bibnamefont
  {Brauner}},\ }\href {\doibase 10.1007/BF02940600} {\bibfield  {journal}
  {\bibinfo  {journal} {Abh.Math.Semin.Univ.Hambg.}\ }\textbf {\bibinfo
  {volume} {6}},\ \bibinfo {pages} {1} (\bibinfo {year} {1928})}\BibitemShut
  {NoStop}%
\bibitem [{\citenamefont {Perron}(1982)}]{perron_noeud_1982}%
  \BibitemOpen
  \bibfield  {author} {\bibinfo {author} {\bibfnamefont {B.}~\bibnamefont
  {Perron}},\ }\href {\doibase 10.1007/BF01396628} {\bibfield  {journal}
  {\bibinfo  {journal} {Inv. Math.}\ }\textbf {\bibinfo {volume} {65}},\
  \bibinfo {pages} {441} (\bibinfo {year} {1982})}\BibitemShut {NoStop}%
\bibitem [{\citenamefont {Taylor}\ and\ \citenamefont
  {Dennis}(2014)}]{taylor_geometry_2014}%
  \BibitemOpen
  \bibfield  {author} {\bibinfo {author} {\bibfnamefont {A.~J.}\ \bibnamefont
  {Taylor}}\ and\ \bibinfo {author} {\bibfnamefont {M.~R.}\ \bibnamefont
  {Dennis}},\ }\href {\doibase 10.1088/1751-8113/47/46/465101} {\bibfield
  {journal} {\bibinfo  {journal} {J. Phys. A: Math. Theor.}\ }\textbf {\bibinfo
  {volume} {47}},\ \bibinfo {pages} {465101} (\bibinfo {year}
  {2014})}\BibitemShut {NoStop}%
\bibitem [{\citenamefont {Lyons}(2003)}]{lyons_elementary_2003}%
  \BibitemOpen
  \bibfield  {author} {\bibinfo {author} {\bibfnamefont {D.~W.}\ \bibnamefont
  {Lyons}},\ }\href {\doibase 10.2307/3219300} {\bibfield  {journal} {\bibinfo
  {journal} {Math. Magazine}\ }\textbf {\bibinfo {volume} {76}},\ \bibinfo
  {pages} {87} (\bibinfo {year} {2003})}\BibitemShut {NoStop}%
\bibitem [{\citenamefont {Urbantke}(2003)}]{urbantke_hopf_2003}%
  \BibitemOpen
  \bibfield  {author} {\bibinfo {author} {\bibfnamefont {H.~K.}\ \bibnamefont
  {Urbantke}},\ }\href {\doibase 10.1016/S0393-0440(02)00121-3} {\bibfield
  {journal} {\bibinfo  {journal} {J. Geom. Phys.}\ }\textbf {\bibinfo {volume}
  {46}},\ \bibinfo {pages} {125} (\bibinfo {year} {2003})}\BibitemShut
  {NoStop}%
\bibitem [{\citenamefont {Mosseri}\ and\ \citenamefont
  {Sadoc}(2012)}]{mosseri_hopf_2012}%
  \BibitemOpen
  \bibfield  {author} {\bibinfo {author} {\bibfnamefont {R.}~\bibnamefont
  {Mosseri}}\ and\ \bibinfo {author} {\bibfnamefont {J.-F.}\ \bibnamefont
  {Sadoc}},\ }\href {\doibase 10.1007/s11224-012-0010-6} {\bibfield  {journal}
  {\bibinfo  {journal} {Struct. Chem.}\ }\textbf {\bibinfo {volume} {23}},\
  \bibinfo {pages} {1071} (\bibinfo {year} {2012})}\BibitemShut {NoStop}%
\bibitem [{\citenamefont {Sadoc}\ and\ \citenamefont
  {Charvolin}(2009)}]{sadoc_3-sphere_2009}%
  \BibitemOpen
  \bibfield  {author} {\bibinfo {author} {\bibfnamefont {J.~F.}\ \bibnamefont
  {Sadoc}}\ and\ \bibinfo {author} {\bibfnamefont {J.}~\bibnamefont
  {Charvolin}},\ }\href {\doibase 10.1088/1751-8113/42/46/465209} {\bibfield
  {journal} {\bibinfo  {journal} {J. Phys. A: Math. Theor.}\ }\textbf {\bibinfo
  {volume} {42}},\ \bibinfo {pages} {465209} (\bibinfo {year}
  {2009})}\BibitemShut {NoStop}%
\bibitem [{\citenamefont {King}(2010)}]{king_knotting_2010}%
  \BibitemOpen
  \bibfield  {author} {\bibinfo {author} {\bibfnamefont {R.~P.}\ \bibnamefont
  {King}},\ }\emph {\bibinfo {title} {Knotting of optical vortices}},\ \href
  {http://eprints.soton.ac.uk/197297/} {\bibinfo {type} {Ph.{D}. {Thesis}}},\
  \bibinfo  {school} {University of Southampton} (\bibinfo {year}
  {2010})\BibitemShut {NoStop}%
\bibitem [{\citenamefont {Houghton}\ \emph {et~al.}(1998)\citenamefont
  {Houghton}, \citenamefont {Manton},\ and\ \citenamefont
  {Sutcliffe}}]{houghton_rational_1998}%
  \BibitemOpen
  \bibfield  {author} {\bibinfo {author} {\bibfnamefont {C.~J.}\ \bibnamefont
  {Houghton}}, \bibinfo {author} {\bibfnamefont {N.~S.}\ \bibnamefont
  {Manton}}, \ and\ \bibinfo {author} {\bibfnamefont {P.~M.}\ \bibnamefont
  {Sutcliffe}},\ }\href {\doibase 10.1016/S0550-3213(97)00619-6} {\bibfield
  {journal} {\bibinfo  {journal} {Nucl. Phys. B}\ }\textbf {\bibinfo {volume}
  {510}},\ \bibinfo {pages} {507} (\bibinfo {year} {1998})}\BibitemShut
  {NoStop}%
\bibitem [{\citenamefont {Stern}(1967)}]{stern_geomagnetic_1967}%
  \BibitemOpen
  \bibfield  {author} {\bibinfo {author} {\bibfnamefont {D.}~\bibnamefont
  {Stern}},\ }\href {\doibase 10.1029/JZ072i015p03995} {\bibfield  {journal}
  {\bibinfo  {journal} {J. Geophys. Res.}\ }\textbf {\bibinfo {volume} {72}},\
  \bibinfo {pages} {3995} (\bibinfo {year} {1967})}\BibitemShut {NoStop}%
\bibitem [{\citenamefont {Stern}(1970)}]{stern_euler_1970}%
  \BibitemOpen
  \bibfield  {author} {\bibinfo {author} {\bibfnamefont {D.~P.}\ \bibnamefont
  {Stern}},\ }\href {\doibase 10.1119/1.1976373} {\bibfield  {journal}
  {\bibinfo  {journal} {American Journal of Physics}\ }\textbf {\bibinfo
  {volume} {38}},\ \bibinfo {pages} {494} (\bibinfo {year} {1970})}\BibitemShut
  {NoStop}%
\bibitem [{\citenamefont {Hesse}\ and\ \citenamefont
  {Schindler}(1988)}]{hesse_theoretical_1988}%
  \BibitemOpen
  \bibfield  {author} {\bibinfo {author} {\bibfnamefont {M.}~\bibnamefont
  {Hesse}}\ and\ \bibinfo {author} {\bibfnamefont {K.}~\bibnamefont
  {Schindler}},\ }\href {\doibase 10.1029/JA093iA06p05559} {\bibfield
  {journal} {\bibinfo  {journal} {J. Geophys. Res.}\ }\textbf {\bibinfo
  {volume} {93}},\ \bibinfo {pages} {5559} (\bibinfo {year}
  {1988})}\BibitemShut {NoStop}%
\bibitem [{\citenamefont {Khurana}(1997)}]{khurana_euler_1997}%
  \BibitemOpen
  \bibfield  {author} {\bibinfo {author} {\bibfnamefont {K.~K.}\ \bibnamefont
  {Khurana}},\ }\href {\doibase 10.1029/97JA00563} {\bibfield  {journal}
  {\bibinfo  {journal} {J. Geophys. Res.}\ }\textbf {\bibinfo {volume} {102}},\
  \bibinfo {pages} {11295} (\bibinfo {year} {1997})}\BibitemShut {NoStop}%
\bibitem [{\citenamefont {Rosner}\ \emph {et~al.}(1989)\citenamefont {Rosner},
  \citenamefont {Low}, \citenamefont {Tsinganos},\ and\ \citenamefont
  {Berger}}]{rosner_relationship_1989}%
  \BibitemOpen
  \bibfield  {author} {\bibinfo {author} {\bibfnamefont {R.}~\bibnamefont
  {Rosner}}, \bibinfo {author} {\bibfnamefont {B.~C.}\ \bibnamefont {Low}},
  \bibinfo {author} {\bibfnamefont {K.}~\bibnamefont {Tsinganos}}, \ and\
  \bibinfo {author} {\bibfnamefont {M.~A.}\ \bibnamefont {Berger}},\ }\href
  {\doibase 10.1080/03091928908218532} {\bibfield  {journal} {\bibinfo
  {journal} {Geophys. Astrophys. Fluid Dyn.}\ }\textbf {\bibinfo {volume}
  {48}},\ \bibinfo {pages} {251} (\bibinfo {year} {1989})}\BibitemShut
  {NoStop}%
\bibitem [{\citenamefont {Whitehead}(1947)}]{whitehead_expression_1947}%
  \BibitemOpen
  \bibfield  {author} {\bibinfo {author} {\bibfnamefont {J.~H.~C.}\
  \bibnamefont {Whitehead}},\ }\href {http://www.jstor.org/stable/87688}
  {\bibfield  {journal} {\bibinfo  {journal} {PNAS}\ }\textbf {\bibinfo
  {volume} {33}},\ \bibinfo {pages} {117} (\bibinfo {year} {1947})}\BibitemShut
  {NoStop}%
\bibitem [{Note1()}]{Note1}%
  \BibitemOpen
  \bibinfo {note} {Mathematically, $\psi (u,v)$ takes its value on the complex
  projective plane, $\protect \mathbb {CP}^1$ (i.e.~the complex numbers with
  the point at $\infty $), which is homeomorphic to the 2-sphere $S^2$: in this
  sense, helicity can be understood as the topological degree of the map $S^3
  \to S^2$.}\BibitemShut {Stop}%
\bibitem [{\citenamefont {Berger}\ and\ \citenamefont
  {Field}(1984)}]{berger_topological_1984}%
  \BibitemOpen
  \bibfield  {author} {\bibinfo {author} {\bibfnamefont {M.~A.}\ \bibnamefont
  {Berger}}\ and\ \bibinfo {author} {\bibfnamefont {G.~B.}\ \bibnamefont
  {Field}},\ }\href {\doibase 10.1017/S0022112084002019} {\bibfield  {journal}
  {\bibinfo  {journal} {J. Fluid Mech.}\ }\textbf {\bibinfo {volume} {147}},\
  \bibinfo {pages} {133} (\bibinfo {year} {1984})}\BibitemShut {NoStop}%
\bibitem [{\citenamefont {Ward}(1999)}]{ward_hopf_1999}%
  \BibitemOpen
  \bibfield  {author} {\bibinfo {author} {\bibfnamefont {R.~S.}\ \bibnamefont
  {Ward}},\ }\href {\doibase 10.1088/0951-7715/12/2/005} {\bibfield  {journal}
  {\bibinfo  {journal} {Nonlinearity}\ }\textbf {\bibinfo {volume} {12}},\
  \bibinfo {pages} {241} (\bibinfo {year} {1999})}\BibitemShut {NoStop}%
\end{thebibliography}%

\end{document}